\documentclass[lettersize, journal]{IEEEtran}
\usepackage[hidelinks]{hyperref}
\usepackage{amssymb,amsfonts}
\usepackage{mathtools}
\usepackage{graphicx}
\usepackage{xpatch}
\usepackage{dblfloatfix}    

\usepackage[caption=false, font=footnotesize]{subfig}

\usepackage[utf8]{inputenc}

\usepackage{sansmath}
\usepackage[defaultmathsizes, italic]{mathastext}

\usepackage{ifdraft}
\usepackage{etoolbox}
\usepackage{mfirstuc}
\usepackage{acro}
\acsetup{single=true}

\NewAcroTemplate{long-short-nop}{\acrowrite{long} \acrowrite{short}}

\DeclareAcronym{soc}{short = SoC, long = system on chip}
\DeclareAcronym{fpga}{short = FPGA, long = field-programmable gate array}
\DeclareAcronym{tpu}{short = TPU, long = tensor processing unit}
\DeclareAcronym{ate}{short = ATE, long = automatic test equipment, short-indefinite=an, long-indefinite=an}
\DeclareAcronym{fom}{short = FOM, long = figure of merit, long-plural-form = figures of merit}
\DeclareAcronym{bist}{short = BIST, long = built-in-self-test}
\DeclareAcronym{pdk}{short = PDK, long = process design kit}

\DeclareAcronym{dim}{short = \textit{d}, long = dimension, first-style=long-short-nop}
\DeclareAcronym{hd}{short=HD, long=hyperdimensional, short-indefinite=an, long-indefinite=a}
\DeclareAcronym{hdc}{short=HDC, long=hyperdimensional computing, short-indefinite=an}
\DeclareAcronym{hdv}{short=hypervector, long=hypervector, first-style=short}
\DeclareAcronym{im}{short=IM, long=item memory, short-indefinite=an, long-indefinite=an}
\DeclareAcronym{am}{short=AM, long=associate memory, short-indefinite=an, long-indefinite=an}
\DeclareAcronym{vsa}{short=VSA, long=vector symbolic architecture}

\DeclareAcronym{ml}{short = ML, long = machine learning, short-indefinite=an}
\DeclareAcronym{dl}{short = DL, long = deep learning}
\DeclareAcronym{nn}{short = NN, long = neural network, short-indefinite=an}
\DeclareAcronym{dnn}{short = DNN, long = deep \ac{nn}}
\DeclareAcronym{cnn}{short = CNN, long = convolutional neural network}
\DeclareAcronym{svm}{short = SVM, long = support vector machine, short-indefinite=an, long-indefinite=a}
\DeclareAcronym{rf}{short = RF, long = random forest, short-indefinite=an}
\DeclareAcronym{mlp}{short = MLP, long = multilayer perceptron, short-indefinite=an}
\DeclareAcronym{knn}{short = KNN, long = k-nearest neighbors algorithm}
\DeclareAcronym{gb}{short = GB, long = gradient boosting}
\DeclareAcronym{pca}{short = PCA, long = principal component analysis}
\DeclareAcronym{tsne}{short = t-SNE, long = t-distributed stochastic neighbor embedding}
\DeclareAcronym{dwt}{short = DWT, long = discrete wavelet transform}

\DeclareAcronym{mse}{short = MSE, long = mean squared error, first-style=long-short, short-indefinite=an}

\DeclareAcronym{dvfs}{short = DVFS, long = Dynamic voltage and frequency scaling}

\DeclareAcronym{pcm}{short = PCM, long = phase-change memory}
\DeclareAcronym{fefet}{short = FeFET, long = Ferroelectric Field-Effect Transistor}

\usepackage{tikz}
\usepackage{circledsteps}

\usepackage[noabbrev,capitalise]{cleveref}
\Crefname{subsection}{Section}{Sections}
\Crefname{subsubsection}{Section}{Sections}
\Crefname{paragraph}{Section}{Sections}
\Crefname{figure}{Fig.}{Fig.}
\Crefname{table}{Tab.}{Tab.}

\usepackage[per-mode=symbol,detect-all, binary-units, list-final-separator={, and }]{siunitx}
\usepackage{microtype}
\usepackage[normalem]{ulem}
\usepackage{tabularx}
\usepackage{booktabs}
\usepackage{multirow}
\usepackage{makecell}
\usepackage{textcomp}

\usepackage{cite}

\usepackage{enumitem}

\makeatletter
\g@addto@macro\@floatboxreset{\centering}
\makeatother

\begin{document}

\title{Brain-Inspired Hyperdimensional Computing: \\ How Thermal-Friendly for Edge Computing?}

\author{Paul R. Genssler, Austin Vas, and Hussam Amrouch
\thanks{Paul R. Genssler, Austin Vas, and Hussam Amrouch are with the Chair of Semiconductor Test and Reliability (STAR), University of Stuttgart, Stuttgart 70569, Germany. E-mail: \{genssler, amrouch\}@iti.uni-stuttgart.de.}%
}

\IEEEoverridecommandlockouts
\IEEEpubid{\begin{minipage}{1.1\textwidth}\ \\[25pt]This work has been submitted to the IEEE for possible publication. Copyright may be transferred without notice, after which this version may no longer be accessible.\end{minipage}}

\maketitle

\begin{abstract}
Brain-inspired \ac{hdc} is an emerging \ac{ml} methods.
It is based on large vectors of binary or bipolar symbols and a few simple mathematical operations.
The promise of \ac{hdc} is a highly efficient implementation for embedded systems like wearables.
While fast implementations have been presented, other constraints have not been considered for edge computing.
In this work, we aim at answering how thermal-friendly HDC for edge computing is.
Devices like smartwatches, smart glasses, or even mobile systems have a restrictive cooling budget due to their limited volume.
Although \ac{hdc} operations are simple, the vectors are large, resulting in a high number of CPU operations and thus a heavy load on the entire system potentially causing temperature violations.
In this work, the impact of \ac{hdc} on the chip's temperature is investigated for the first time. 
We measure the temperature and power consumption of a commercial embedded system and compare \ac{hdc} with conventional CNN.
We reveal that \ac{hdc} causes up to \SI{6.8}{\degreeCelsius} higher temperatures and leads to up to \SI{47}{\percent} more CPU throttling.
Even when both \ac{hdc} and CNN aim for the same throughput (i.e., perform a similar number of classifications per second), \ac{hdc} still causes higher on-chip temperatures due to the larger power consumption.
\end{abstract}

\acresetall


\section{Introduction}
\label{sec intro}

\IEEEPARstart{W}{earable} devices are an emergent type of embedded system. 
Smartwatches, fitness trackers, e-textiles, or smart glasses are prominent examples.
They detect seizures, measure the heart rate, categorize sport activities, augment reality to help the user, or enable wireless payments, among others.
Developers of such devices are trending towards embed\-ded intelligence for more independence from phones and also for immediate feedback.
The required \ac{ml} and \ac{dl} methods typically involve heavy computations and require relatively large storage for the models. 
However, in embedded systems, resources such as processing power, memory, energy, and cooling are tightly constrained.

Brain-inspired \ac{hdc} has emerged as a promising alternative to the established \ac{ml} methods.
Similar or better inference accuracy compared to deep \acp{nn} has been demonstrated for speech recognition \cite{voicehdhyperdimensionalcomputing2017imani}, EEG-based seizure detection \cite{ensemblehyperdimensionalclassifiers2020burrello}, EMG-based gesture detection \cite{wearablebiosensingsystem2021moin}, or wafer map defect pattern classification \cite{braininspiredcomputingwafer2021genssler}, among others \cite{classificationusinghyperdimensional2020ge}.
While layers of neurons are the base in \acp{nn}, large vectors with dimensions in the thousands, \acp{hdv}, form the basic blocks in \ac{hdc}.
Their components can be simple bits, bipolar values, or more complex numbers and they are all independent of each other.
Hence, parallelization is easy and consequently, computationally efficient implementations have been proposed for ASICs \cite{pulphdacceleratingbraininspired2018montagna}, FPGAs \cite{hardwareoptimizationsdense2019schmuck}, and embedded devices \cite{ensemblehyperdimensionalclassifiers2020burrello, wearablebiosensingsystem2021moin}.

Three basic operations like dot product, component-wise majority, and multiplication are used to encode the real-world data into non-binary \acp{hdv}.
Compared to the floating-point matrix multiplications required by \acp{nn}, such operations are lightweight, i.e., computationally less expensive.
Therefore, \ac{hdc} has been proposed as a promising \ac{ml} method for embedded systems and smart things \cite{voicehdhyperdimensionalcomputing2017imani, ensemblehyperdimensionalclassifiers2020burrello, wearablebiosensingsystem2021moin, classificationusinghyperdimensional2020ge, pulphdacceleratingbraininspired2018montagna, hardwareoptimizationsdense2019schmuck}.
Other advantages include the robustness against noise in input data, e.g., from unreliable cheap sensors, or in the model's memory as well as the ability to learn from very little data \cite{ensemblehyperdimensionalclassifiers2020burrello}.
To realize the fast product cycle demanded by consumers, off-the-shelf hardware and software implementations are preferred by developers for their short time to market. 
Highly specialized hardware like ASICs achieve better computational performance and higher energy efficiency.
Nevertheless, \ac{hdc} remains a demanding application due to the large size of the \acp{hdv}, data movement, and complex encoding.

In addition to memory, time, and power constraints face embedded systems, wearables in particular, also a temperature constraint.
Such systems cannot be actively cooled because fans draw additional power, contain error-prone moving parts, and require space that is not available, e.g., in smart glasses.
Passive cooling is limited on one hand by the overall size of the system and thus its thermal capacity.
On the other hand, if and how it has contact with human skin can drastically limit the maximum sustainable temperature.
\textit{The impact of \ac{hdc} on the on-chip temperatures has not yet been evaluated.}

Our main contribution: We are the first to investigate if \ac{hdc} conforms with the thermal constraint that an embedded system imposes.
If it causes temperature violations due to the demanding operations, then the promise as an alternative \ac{ml} method for off-the-shelf embedded systems needs to be revisited.
We employ an optimized state-of-the-art implementation of \ac{hdc} \cite{onlinehdrobustefficient2021hernandez-cane} on a commercial Raspberry Pi 4 to classify images from the Fashion-MNIST data set \cite{fashionMNIST2017xiao}, a common task, e.g., in smart glasses.
We measure the temperature of the CPU and the power consumption of the whole system under realistic workloads.
\Ac{hdc} is compared against \iac{cnn} during training and inference.
We reveal that only \ac{hdc} causes CPU throttling due to higher temperatures but still trains faster.
In a scenario where both methods perform equally fast, \ac{hdc} consumes more energy and causes higher on-chip temperatures.

\section{Preliminaries}
\label{sec background}

The concepts of brain-inspired \ac{hdc} can be implemented with different underlying \acp{vsa}, e.g., with vectors of simple bits, bipolar values, integers, or real numbers. 
Depending on the type, different mathematical operations realize the basic operations of binding, bundling, permutation, and similarity score.
In this work, real number \acp{hdv} are employed because they offer the highest inference accuracy.
A drawback is their higher computational cost compared to binary implementations.

To improve the inference accuracy, the concept of retraining has been proposed \cite{voicehdhyperdimensionalcomputing2017imani}. 
After an initial training cycle and class \ac{hdv} generation, the training data itself is used for inference.
If a sample is incorrectly classified, then the class \ac{hdv} is adjusted to be more similar to the sample. 
The process is repeated.
Each iteration is referred to as an epoch.

Temperature can be a limiting factor in embedded devices.
It correlates with the amount of processing performed by the compute units, which can be CPUs, GPUs, or other specialized ones, like \acp{tpu}. 
In addition, memory accesses consume comparatively large amounts of electric energy, which translates again into thermal energy. 
Both sources of heat have to be controlled to achieve an overall lower system-wide temperature. 
\Ac{dvfs} is one of the most important tools to manage the temperature. 
By reducing the clock frequency, the voltage can be reduced, having a quadratic impact on the dynamic power consumption. 

\section{Experimental Setup}
\label{sec setup}

\begin{figure}
    \centering
    \includegraphics{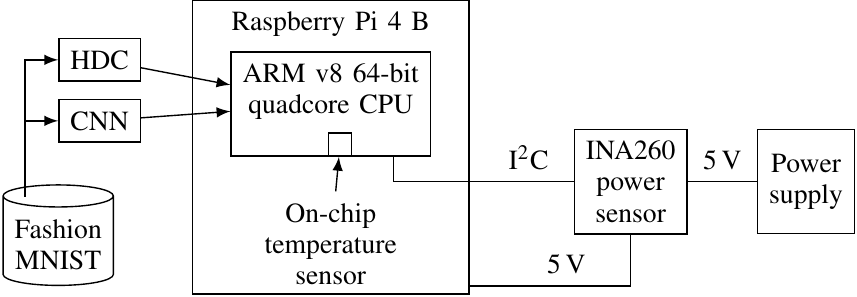}
    \caption{Experimental setup with Fashion MNIST as a realistic workload.}
    \label{fig setup}
\end{figure}

A Raspberry Pi\,4 Model\,B with \SI{8}{\giga\byte} RAM is employed in the experiments.
The Raspberry Pi OS (32-bit Linux) uses kernel version 5.10.17-v7l and the firmware from May 27, 2021.
The nominal CPU frequency is \SI{1500}{\mega\hertz} and starts throttling once its temperature exceeds \SI{80}{\degreeCelsius} consistently (``soft max'').
In the experiments, this throttling prevents the temperature from reaching the maximum of \SI{85}{\degreeCelsius}.
Temperature is measured with the on-chip sensor of the Broadcom 2711 \ac{soc} and at idle is on average \SI{45}{\degreeCelsius}.
The ambient temperature is kept constant at about \SI{20}{\degreeCelsius} in all experiments for fair comparisons.
The board is not covered, in a case, nor is a heat sink attached.
Power measurements are done with an intermediary INA260 sensor between the USB power adapter output and the board, as shown in \cref{fig setup}.
The measurements are collected by the Pi itself via I\textsuperscript{2}C. 
Idle power consumption is about \SI{3.5}{\watt} with the CPU frequency throttled down to \SI{600}{\mega\hertz} so save power.

Image classification is a demanding task for embedded systems.
Each sample contains a lot of data that has to be processed. 
In this work, the Fashion-MNIST data set is employed as an example \cite{fashionMNIST2017xiao}.
It comprises \num{60000} training and \num{10000} testing samples of ten different classes of clothing.
All images are grayscale and have a resolution of 28x28 pixels.

OnlineHD \cite{onlinehdrobustefficient2021hernandez-cane} is a state-of-the-art \ac{hdc} implementation using bipolar \acp{hdv}. 
Version 0.1.2 is built on top of Python's PyTorch framework and provides efficient implementations for the underlying mathematical operations during encoding.
The similarity computation is written in C++.

Inspired by the visual mechanism in the brain, \acp{cnn} were proposed as a special form of a neural network with convolution layers, sub-sampling layers, and completely connected layers.
Two \ac{cnn} architectures are described in \cref{tab cnn arch} and implemented with TensorFlow 1.14.0 and Python 3.7.3.
The larger \ac{cnn} is used in the experiments, while the simplified CNN is discussed in \cref{sec results simple cnn}.
Neither model utilizes the GPU during training or inference, all processing is done on the CPU.

\begin{table}
    \centering 
    \caption{\acs*{cnn} architectures employed in the experiments.}
    \label{tab cnn arch}
    \begin{tabular}{lrS[table-format=6]r}
    	\toprule
    	Layer type   & Output shape                        & {Parameters}          & Simplified CNN        \\ \midrule
    	Conv2D       & (28, 28, 64)                        & 320                   & (22, 22, 6)           \\
    	MaxPooling2D & (14, 14, 64)                        & 0                     & (11, 11, 6)           \\
    	Dropout      & (14, 14, 64)                        & 0                     & (11, 11, 6)           \\
    	Conv2D       & (14, 14, 32)                        & 8224                  & removed               \\
    	MaxPooling2D & (7, 7, 32)                          & 0                     & removed               \\
    	Dropout      & (7, 7, 32)                          & 0                     & removed               \\
    	Flatten      & (1568)                              & 0                     & (726)                 \\
    	Dense        & (256)                               & 401664                & (20)                  \\
    	Dropout      & (256)                               & 0                     & (20)                  \\
    	Dense        & (10)                                & 2570                  & (10)                  \\ \midrule
    	\multicolumn{2}{l}{Total trainable parameters}     & 412778                & 15050                 \\
    	\multicolumn{2}{l}{Inference accuracy (70 epochs)} & {\SI{92.7}{\percent}} & \SI{85.8}{\percent} \\ \bottomrule
    \end{tabular} 
\end{table}

\section{Experimental Results}
\label{sec results}

All tasks, such as training and inference, are continuously repeated for four hours.
This allows the system to accumulate heat and to minimize environmental influences.
The results presented in the figures are short representative segments from these long experiments.
The reported numbers are averaged over the whole experiment.

\begin{table}
    \centering
    \caption{Inference accuracy for different dimensions with \acs*{hdc} and a nine degree polynomial SVM for comparison.}
    \label{tab dimension}
    \begin{tabular}{l*{5}{r}}
    	\toprule
    	Dimension          & \phantom{00}\num{500} & \phantom{0}\num{2000} & \phantom{0}\num{4000} &  \num{10000} & SVM \\ \midrule
    	ISOLET &                  92.2 &                  93.2 &                  93.4 &         93.6 & 96.3 \\
    	MNIST  &                  91.2 &                  94.4 &                  94.8 &         95.1 & 98.4 \\
    	FMNIST             &                  82.3 &                  84.7 &                  86.0 &         86.7 & 89.4 \\
    	\bottomrule
    \end{tabular}
\end{table}

\subsection{Inference Accuracy}

Each model, \ac{cnn} and \ac{hdc}, is trained on a server machine with no memory, temperature, or power constraint. 
Training the models longer does not increase the energy consumption or the thermal emissions of the deployed model. 
During training, only the neurons' parameters or the \acp{hdv} are changed, not the \ac{cnn} architecture itself or the dimension.
Therefore, fully trained models can be deployed to the target system.
After training for \num{70} epochs, the \ac{cnn} achieves an inference accuracy of \SI{92.7}{\percent} on the test set.
For \ac{hdc}, the various dimensions are explored and summarized in \cref{tab dimension}, from \num{500} up to \num{10000} achieving an inference accuracy of \SI{82.3}{\percent} and \SI{86.7}{\percent}, respectively.
Even though this accuracy increases with the dimension, anything above \num{4000} (\SI{86.0}{\percent}) would have a diminishing return for an embedded system.
It increases energy consumption and inference latency.
Hence, a dimension of \num{4000} is selected for the remaining experiments.

\subsection{Training Analysis on the Embedded System}

\begin{figure}
    \centering
    \includegraphics{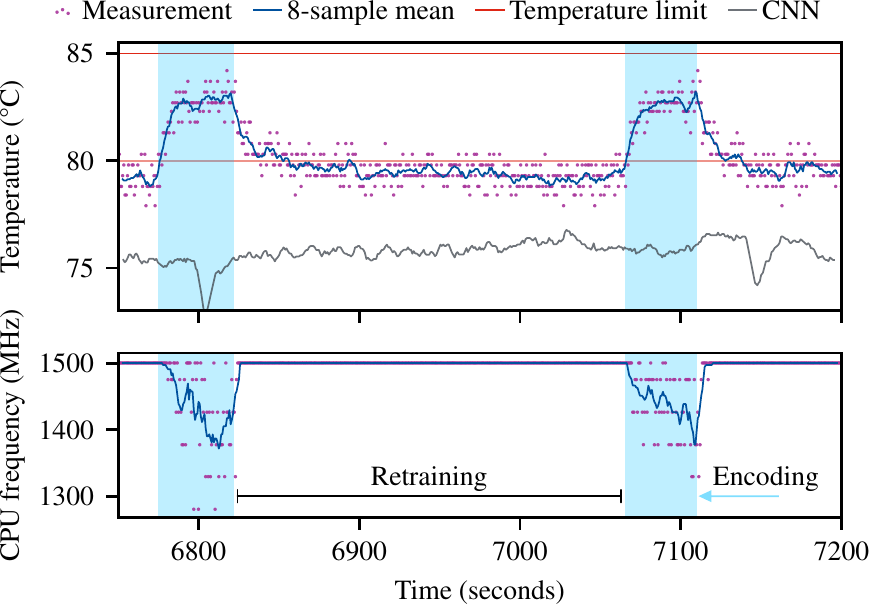}
    \caption{Encoding during repeated training in \acs*{hdc} causes CPU throttling. Despite that, \acs*{hdc} is still faster in training the model.}
    \label{fig train temp}
\end{figure}

The training of an \ac{ml} model involves lots of data and computations.
Hence, it is usually performed in the cloud and only the fully trained model is deployed to the target system.
However, for personalized applications, like seizure detection or voice recognition, the model could be trained further to improve inference performance with the user.
In this experiment, the model was trained from scratch to evaluate the overall impact of this type of workload on temperature. 
The data set is loaded once into memory and the training is repeated to simulate a larger data set.
Re-/training epochs are reduced to one and 20, resulting in a reduced inference accuracy of \SI{86.0}{\percent} and \SI{81.9}{\percent} for the \ac{cnn} and \ac{hdc}, respectively.
The \ac{cnn} does not exceed \SI{79}{\degreeCelsius} with \SI{75.6}{\degreeCelsius} on average as shown in \cref{fig train temp}.
In contrast, \ac{hdc} reaches temperatures of up to \SI{84.2}{\degreeCelsius}, \SI{6.7}{\degreeCelsius} more than \ac{cnn} on average.
\Ac{hdc} is frequently close to the maximum temperature of \SI{85}{\degreeCelsius} during the encoding of the data set. 
Hence, \ac{dvfs} is employed by the \ac{soc}, reducing the clock frequency up to \SI{47}{\percent} (\SI{800}{\mega\hertz}). 
During retraining, the temperature drops to about \SI{79}{\degreeCelsius}.
No thermal buffers are created for the next encoding cycle.
The power consumption is similar for both models with \SI{5.2}{\watt} on average.
Therefore, the operations used by \ac{hdc} cause more heat compared to the \ac{cnn}.
Yet, despite the CPU throttling, \ac{hdc} completes a training cycle \SI{14.9}{\percent} faster as listed in \cref{tab images per second}.
However, inference accuracy on the test set is \SI{4}{\percent} lower with \ac{hdc}.
            
\begin{table}
    \centering 
    \caption{Average number of images processed per second.}
    \label{tab images per second}
    \begin{tabular}{c*{3}{S[table-format=4.1]}}
        \toprule
        Scenario  & {Training} & {Inference} & {Inference} \\
        Temperature Limit & {\SI{85}{\degreeCelsius}} & {\SI{85}{\degreeCelsius}} & {\SI{60}{\degreeCelsius}} \\
        \midrule
        \acs*{hdc} & 206.3  & 1000.5 & 382.0 \\
        \acs*{cnn}  & 179.5  &  322.8 & 373.0  \\
        \bottomrule
    \end{tabular}
\end{table}

\subsection{Continuous Inference Analysis}
\label{sec results cont inf}
\label{sec results simple cnn}

The primary task of \iac{ml} model in an embedded system is the inference of incoming samples. 
Smart glasses could analyze the live camera feed to overlay directions for an augmented reality experience.
A smartphone listens for keywords to activate the voice assistant.
To mimic a similar scenario, the inference of the test data set is continuously repeated.

\Ac{hdc} processes \num{1000.5} images per second, which is 3.1x more than \ac{cnn} because of its less complex operations.
It also utilizes the CPU to its full extend, whereas the CPU load during \ac{cnn} inference is only about \SI{75}{\percent}.
However, because of the more complex operations, the \ac{cnn} consumes \SI{5.87}{\watt} compared to \ac{hdc} with \SI{5.44}{\watt}, on average.
Although power consumption is \SI{8}{\percent} higher with a \ac{cnn}, the temperature is \SI{2.1}{\degreeCelsius} lower on average. 
Furthermore, the \ac{cnn} exceeds the first temperature limit of \SI{80}{\degreeCelsius} for \SI{7.2}{\percent} of the total time compared to \SI{47.5}{\percent} with \ac{hdc}.
The trend in the data for \ac{cnn} suggests an increase in temperature above \SI{80}{\degreeCelsius} if the data set was larger.
However, the setup time between two inference runs reduces CPU load and temperature. 
During this setup period, book-keeping tasks are done and the inference accuracy calculated.
The same applies to \ac{hdc}, although the temperature increases faster and could reach \SI{85}{\degreeCelsius}, triggering \ac{dvfs}.
The measured CPU clock frequency reduction is not because of over-temperature but enabled by the reduced CPU load during setup.

A simplified \ac{cnn} has been tested as well, its structure is described in \cref{tab cnn arch}.
Due to its reduced complexity, the number of processed images increase to \num{2427.5} outperforming \ac{hdc}.
Power consumption remains at the level of \SI{5.3}{\watt}.

\subsection{Inference Analysis under Temperature Constraint}

\begin{figure}
    \centering
    \includegraphics{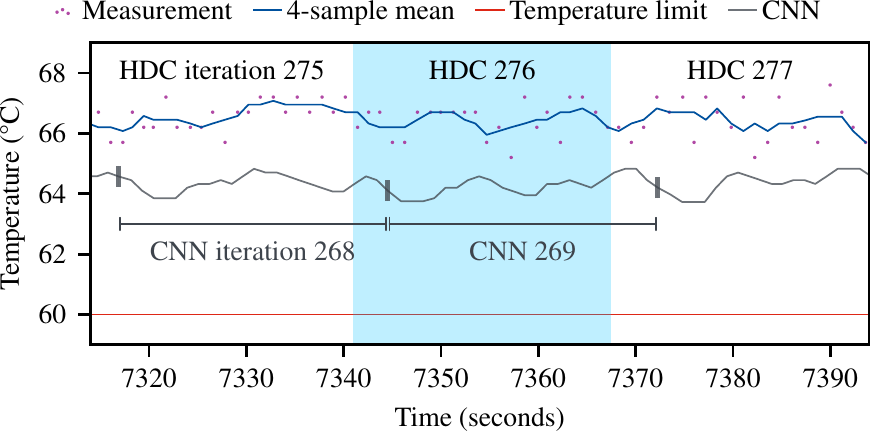}
    \caption{Inference with a temperature constraints of \SI{60}{\degreeCelsius}. Both models cause permanent CPU throttling to \SI{600}{\mega\hertz}. In addition, the temperature with \acs*{hdc} is \SI{2}{\degreeCelsius} higher but both classify a similar number of images per second.}
    \label{fig inf60 results}
\end{figure}

A Raspberry Pi has an upper temperature limit of \SI{85}{\degreeCelsius}, an integrated heat spreader made of metal, and a large PCB to dissipate the thermal energy.
Smaller edge devices, embedded systems, and wearables do not have such high temperature limits or thermal mass. 
To simulate these tighter temperature constraints with a Raspberry Pi, its clock frequency is throttled if a temperature threshold is crossed. 
In this experiment, the threshold is set to \SI{60}{\degreeCelsius}. 
If the devices touch the skin, such high temperatures could harm the user.
A small script checks the temperature every second and sets the CPU frequency to \SI{600}{\mega\hertz} in the case of overheating. 
As shown in \cref{fig inf60 results}, both models exceed the limit even with a throttled CPU clock.
The number of processed images is reduced from \num{1001} (no limit, \cref{sec results cont inf}) to \num{382} for \ac{hdc} and increases for \ac{cnn} from \num{323} to \num{373}.
\Ac{hdc} has a \SI{22}{\percent} higher CPU load and consumes \SI{7}{\percent} more energy.
This results in a \SI{2}{\degreeCelsius} higher temperature on average.

\subsection{Hand-crafted Encoding for ISOLET}
To provide an additional perspective, the ISOLET dataset \cite{isolet} of voice samples is analyzed.
The first model is OnlineHD, which achieves an accuracy of \SI{94.7}{\percent} with a dimension of 2048.
The second model is based on the C-Code implementation of HD-Lib\footnote{\url{https://github.com/skurella/hdlib}}, uses binary vectors of dimension 4096 and a record encoding to achieve an accuracy of \SI{87.1}{\percent} (without retraining). 
The third model is a KNN classifier with $n=9$ and \SI{92.1}{\percent} accuracy implemented with the Scikit-learn library.  

OnlineHD, HD-Lib, and KNN process \SIlist[list-units=single]{2369.5;537.7;390.1}{samples\per\second}, respectively.
OnlineHD also consumes the most energy (\SI{5.9}{\watt}) and generates the most heat (\SI{82.8}{\degreeCelsius}) reducing the clock frequency to \SI{1405}{\mega\hertz} on average. 
The HD-Lib-based hand-crafted encoding consumes slightly less power (\SI{5.7}{\watt}) and consequently generates less heat (\SI{82.1}{\degreeCelsius}) but still throttles the CPU to \SI{1431}{\mega\hertz} on average. 
The KNN does not utilize all CPU cores fully and the average temperature is about \SI{74.9}{\degreeCelsius}, not throttling the CPU and consuming less power (\SI{5.5}{\watt}).

\section{Discussion}

\tikzset{/csteps/inner color=white, /csteps/fill color=black}

\cstep{} \ac{hdc} performs training and inference faster than \ac{cnn}.
However, under temperature constraints and thus reduced CPU frequency because of throttling, the throughput (images per second) is reduced more with \ac{hdc} compared to \ac{cnn}.
The number of images processed per second decreases by \SI{33}{\percent} and \SI{62}{\percent} during training and inference, respectively. 
\cstep{} In contrast, with \ac{cnn}, the throughput decreases \SI{16}{\percent} and even increases \SI{16}{\percent} during training and inference, respectively.
These results suggest \ac{hdc} is \textit{compute-bound} since a reduction in frequency by \SI{60}{\percent} reduces the inference time by \SI{62}{\percent}.
The \ac{cnn}, to the contrary, is \textit{memory-bound}.
The CPU is stalled waiting for data.
A reduced frequency reduces this wait time and the CPU utilization increases.
\cstep{} Nevertheless, \ac{hdc} consistently causes higher temperatures than \ac{cnn}.
During inference, \ac{hdc} exceeded the targeted \SI{80}{\degreeCelsius} almost half of the time, 6.6x more often than \ac{cnn},
yet \ac{hdc} performs 3.1x more inferences per second.
However, even with reduced CPU frequency, \ac{hdc} exceeds the temperature limit by \SI{2}{\degreeCelsius} more than \ac{cnn}, while both have comparable inference speed.
Because of its high number of computations compared to \ac{cnn}, \ac{hdc} performs worse regarding temperature and power consumption in such low-power scenarios expected for wearable devices. 
\cstep{} The efficiency of the \ac{cnn} is improved by using a 64-bit OS.
The number of images processed per second jumps to \num{747} while the power consumption is reduced to \SI{5.0}{\watt} at the same time. 
Alternatively, quantizing the CNN to 8-bit integers doubles the processing speed.
A simplified \ac{cnn}, described in \cref{tab cnn arch}, achieves after 70 epochs training the same accuracy as \ac{hdc} but processes \SI{2427.5}{images\per\second} at \SI{5.3}{\watt}, although at \SI{77.7}{\degreeCelsius}.
Hand-crafted \ac{hdc} encoding on a different dataset suggests a similar pattern.
With off-the-shelf hardware, \ac{hdc} is outperformed by other methods in non-functional metrics, like power consumption and temperature.

\section{Conclusion}
\label{sec conclusion}
This is the first work to investigate the impact of \ac{hdc} on the on-chip temperature of a CPU.
We conducted several experiments on a commercial embedded system.
Power consumption and temperature are measured while executing \ac{hdc} and \ac{cnn} algorithms for both training and inference.
We unveil that \ac{hdc} faces a temperature and power challenge in low-power systems such as wearables and hence it is less thermal-friendly in edge computing.

\section*{Acknowledgment}
This research was supported by Advantest as part of the Graduate School ``Intelligent Methods for Test and Reliability'' (GS-IMTR) at the University of Stuttgart.


\end{document}